\documentclass[12pt]{spieman}  
\usepackage{amsmath,amsfonts,amssymb}
\usepackage{graphicx}
\usepackage{setspace}
\usepackage{tocloft}
\usepackage{lineno}
\usepackage[dvipsnames]{xcolor}

\newcommand{\fe}{$^{55}$Fe}

\title{X-ray spectral performance of the Sony IMX290 CMOS sensor near Fano limit after a per-pixel gain calibration}

\author[a,*]{Benjamin~Schneider}
\author[a]{Gregory~Prigozhin}
\author[a]{Richard~F.~Foster}
\author[a]{Marshall~W.~Bautz}
\author[a]{Hope~Fu}
\author[a]{Catherine~E.~Grant}
\author[a]{Sarah~Heine}
\author[a]{Jill~Juneau}
\author[a]{Beverly~LaMarr}
\author[b]{Olivier~Limousin}
\author[a]{Nathan~Lourie}
\author[a]{Andrew~Malonis}
\author[a]{Eric~D.~Miller}

\affil[a]{Kavli Institute for Astrophysics and Space Research, Massachusetts Institute of Technology, Cambridge, MA, USA}
\affil[b]{Université Paris-Saclay, Université Paris Cité, CEA, CNRS, AIM, 91191 Gif-sur-Yvette, France}

\cftpagenumbersoff{figure}
\cftpagenumbersoff{table} 
\begin{document} 
\maketitle

\begin{abstract}
The advent of back-illuminated complementary metal-oxide-semiconductor (CMOS) sensors and their well-known advantages over charge-coupled devices (CCDs) make them an attractive technology for future X-ray missions. However, numerous challenges remain, including improving their depletion depth and identifying effective methods to calculate per-pixel gain conversion.
We have tested a commercial Sony IMX290LLR CMOS sensor under X-ray light using an \fe\ radioactive source and collected X-ray photons for $\sim$15 consecutive days under stable conditions at regulated temperatures of 21°C and 26°C. 
At each temperature, the data set contained enough X-ray photons to produce one spectrum per pixel consisting only of single-pixel events. We determined the gain dispersion of its 2.1 million pixels using the peak fitting and the Energy Calibration by Correlation (ECC) methods. We measured a gain dispersion of 0.4\% at both temperatures and demonstrated the advantage of the ECC method in the case of spectra with low statistics.
The energy resolution at 5.9~keV after the per-pixel gain correction is improved by $\gtrsim$10~eV for single-pixel and all event spectra, with single-pixel event energy resolution reaching $123.6\pm 0.2$~eV, close to the Fano limit of silicon sensors at room temperature. 
Finally, our long data acquisition demonstrated the excellent stability of the detector over more than 30~days under a flux of $10^4$ photons per second.

\end{abstract}


{\noindent \footnotesize\textbf{*}Benjamin Schneider, \linkable{bschn@mit.edu}}


\section{Introduction}
\label{sect:intro}  
For X-ray astronomy, charge-coupled devices (CCDs) have been successfully used over the past decades \cite{struder2001a,garmire2003a,koyama2007a,predehl2021a,schneider2023a}, and are still proposed for upcoming and future X-ray missions \cite{bautz2020a,lamarr2022a,miller2022a,miller2023a}. CCDs are a mature technology that has benefited from continuous development, achieving very low noise and excellent uniformity.
The next generation of X-ray missions requires higher readout speed, a lower power consumption and a better radiation tolerance while maintaining high quality imaging and spectral performance. Recent improvements of complementary metal oxide semiconductor (CMOS) detectors and the use of back-side illuminated CMOS offer a promising alternative to CCDs for future X-ray instruments. In particular, CMOS detectors provide a high readout frame rate per second (fps), low readout noise, low power consumption, and can be operated at room temperature. In addition, the active pixel design of CMOS prevents the inherent and growing charge transfer inefficiency affecting CCD performance over time due to space radiation. Their relatively low cost compared to CCDs makes them an attractive technology for wide-field X-ray instruments. The wide-field X-ray telescope (WXT), on board the Einstein Probe (EP)\cite{yuan2022a} mission, launched in early 2024, and its pathfinder the Lobster Eye Imager for Astronomy (LEIA) flight experiment\cite{zhang2022a}, launched in 2022, are the first X-ray missions to use scientific CMOS sensors in space environment and have already demonstrated promising results\cite{liu2024a,levan2024a}. 
However, many challenges remain before CMOS sensors can be considered a viable alternative for more X-ray applications. CMOS devices suffer from lower depletion depths and reduced X-ray quantum efficiency compared to CCDs\cite{stefanov2022a}, though manufacturers have been improving X-ray optimized designs\cite{wu2022a,townsend-rose2023a}. The highly parallelized readout architecture of CMOS devices, which provides their high readout rate, also leads to higher variation in gain from pixel-to-pixel compared to that from the single readout node of a CCD array. This gain variation can degrade the spectral energy resolution\cite{wang2019b,narukage2020a,hsiao2022a,wu2022a,townsend-rose2023a}.
Improved procedures to characterize the pixel-to-pixel gain variation inherent to CMOS devices can significantly improve their spectral energy resolution.
One commercial CMOS sensor that has shown promise for detection of X-rays is the Sony IMX290LLR ($1920\times1080 \ 2.9~\mu$m pixels), which is optimized for optical light and routinely used for astrophotography or security camera systems. However, its low noise ($\sim$2 
e- rms) and back-side illuminated design makes it suitable for detecting X-rays. Previous works have successfully tested the device under X-ray light and demonstrated its excellent spectral performance from 250~eV to 6.4~keV \cite{tammes2020a,roth2022a}. 
In this paper, we investigate an efficient technique to characterize per-pixel conversion gain calibration using a CMOS sensor. We measure and correct the per-pixel gains of the 2.1 million pixels composing the Sony IMX290 CMOS sensor. We then derive the 5.9~keV spectral performance of the device after correcting for gain dispersion. We finally report on the stability of the system over time, based on the long acquisition periods required to measure the gain of each pixel.

\section{Experimental setup}
\label{sect:exp_setup}
\begin{figure}[t]
    \centering
    \includegraphics[width=0.6\hsize]{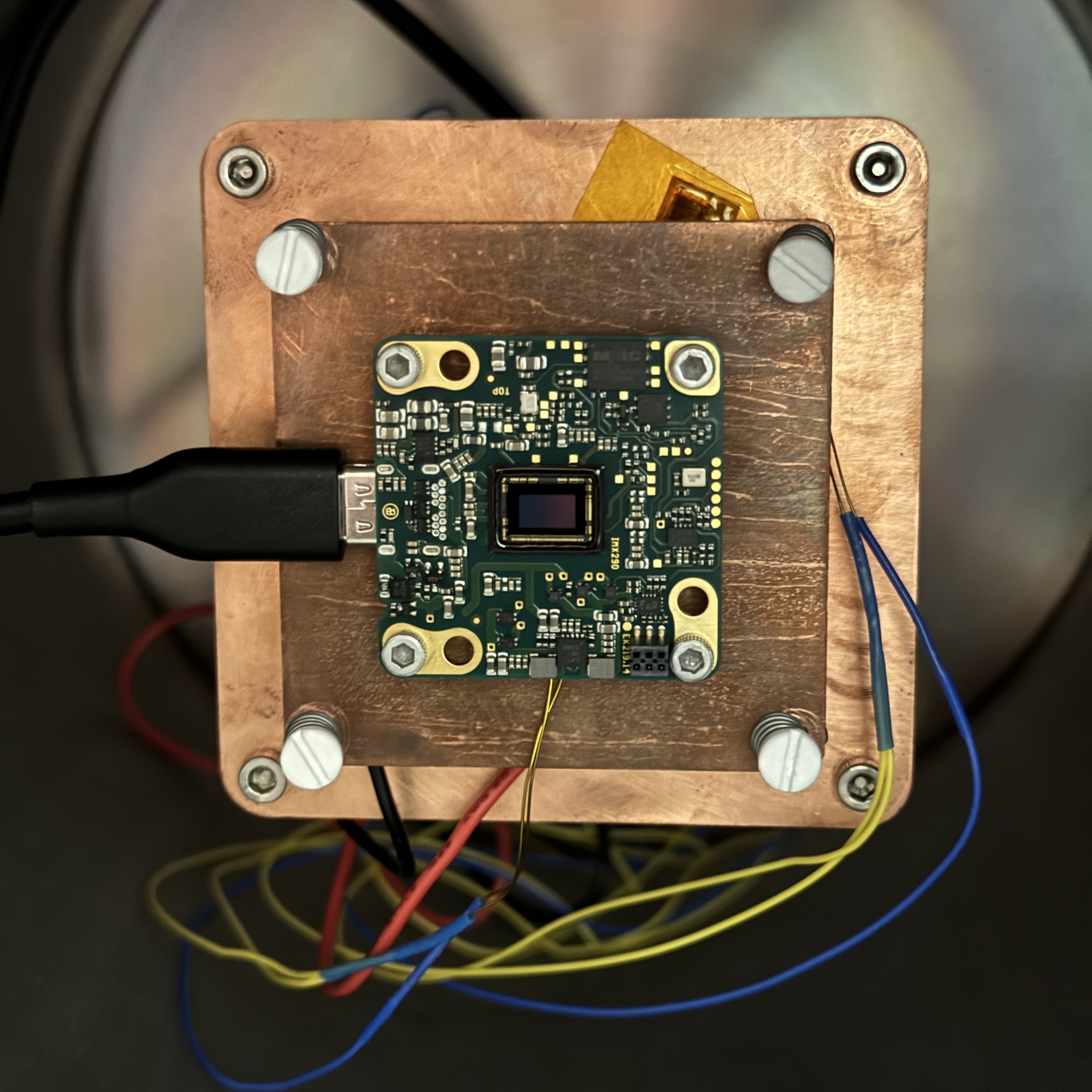}
    \caption 
    { \label{fig:CMOS}
    Picture of the experimental setup. The CMOS sensor is visible at the center of the image and surrounded by its readout electronics. The camera is connected to the cold side of the TEC (not visible) via a copper plate. The hot and cold sides are held together with PTFE plastic screws to minimize the heating load.} 
\end{figure}

Our setup employs the IDS Imaging UI-3862LE-M camera offering a board-level, low-cost, compact and versatile system. The board is equipped with a Sony STARVIS I IMX290 CMOS sensor (see Table~\ref{tab:sensor}), which is a low-noise monochrome back-side illuminated device making it suitable to detect X-ray light. The board uses USB3.1 for command/control, power and image transfer enabling a readout speed up to 120~fps. As is standard for X-ray imaging, the cover glass over the top of the sensor package was removed to ensure the detection of X-ray photons below 10~keV. The camera was installed in a cryostat and operated under vacuum. A thermoelectric cooler (TEC) was used to cool down the camera and a proportional–integral–derivative (PID) controller was implemented to maintain a constant temperature ($\pm$0.2°C) during measurements. The hot and cold sides of the TEC were clamped to copper interfaces to maximize the thermal conduction and were monitored by two resistance temperature detectors (RTDs). The hot side was connected to a power feedthrough to evacuate the heat produced by the camera and TEC in the cryostat. External to the power feedthrough, a cold plate was attached and liquid cooled by a chiller at 10°C, which maintained the hot side at a constant temperature and improved the TEC cooling capacity (see Fig.~\ref{fig:CMOS}).

Preliminary tests illuminated the sensor with a radioactive source of \fe\ producing Mn-K$\alpha$ (5.90~keV) and Mn-K$\beta$ (6.49~keV) emission lines. During that phase, the analog gain of the image sensor was tuned to obtain a conversion gain of 1 e-/ADU and maintain the electronic noise as low as possible. We also illuminated the system with a source composed of $^{210}$Po and Teflon to produce fluorescence lines of C-K (0.53~keV) and F-K (0.68~keV). Both sources validate the ability of this sensor to detect soft X-ray photons down to 277~eV, as previously observed in \citenum{tammes2020a}. Only the results from the \fe\ measurements are presented in this paper.

\begin{table}[ht]
    \caption{Main characteristics of the IDS UI-3862LE camera equipped with the Sony IMX290 CMOS sensor.} 
    \label{tab:sensor}
    \begin{center}       
        \begin{tabular}{|l|l|} 
            \hline
            \rule[-1ex]{0pt}{3.5ex}  Camera model & UI-3862LE-M  \\
            \hline
            \rule[-1ex]{0pt}{3.5ex}  Sensor model & IMX290LLR-C  \\
            \hline
            \rule[-1ex]{0pt}{3.5ex}  Imaging area & 5.610 mm $\times$ 3.175 mm  \\
            \hline
            \rule[-1ex]{0pt}{3.5ex}  Number of pixels & 1936 $\times$ 1096 \\
            \hline
            \rule[-1ex]{0pt}{3.5ex}  Pixel size & 2.9~µm   \\
            \hline
            \rule[-1ex]{0pt}{3.5ex}  Pixel clock range & 20 -- 474 MHz \\
            \hline
            \rule[-1ex]{0pt}{3.5ex}  Frame rate & up to 120~fps \\
            \hline
            \rule[-1ex]{0pt}{3.5ex}  Exposure time & 0.011 ms -- 120 s \\
            \hline
            \rule[-1ex]{0pt}{3.5ex}  Readout noise & 1.9 e- rms  \\
            \hline
            \rule[-1ex]{0pt}{3.5ex}  ADC & 12~bit  \\
            \hline
            \rule[-1ex]{0pt}{3.5ex}  Shutter & Rolling shutter  \\
            \hline
            \rule[-1ex]{0pt}{3.5ex} Camera power &  0.9 -- 1.5 W \\
            \hline 
        \end{tabular}
    \end{center}
\end{table}

\section{Real time data processing}
\label{sect:data_processing}
We used the uEye interface for Python (PyuEye), provided by IDS, to set the camera settings (e.g., fps, pixel clocks, gain) and access full raw frames. The hot pixel correction offered by the API was disabled to prevent single X-ray events from being considered as hot pixels.
The camera was operated at 10 fps, with each frame producing 4.2~MB of data, thus generating $\sim$2.5~GB every minute. For experimental testing, the large amount of data generated raised multiple challenges, especially in terms of storage capacity and post-processing time. In addition, pixels hit by X-ray photons represent only a small fraction of the entire image. To overcome these challenges, we developed an algorithm to process the frames in real time to only extract and save valid X-ray events which drastically reduced the memory usage by only keep the relevant information deposited by X-ray photons in each frame. 

For real-time X-ray event finding, the imaging sequence starts by taking a series of initial frames, of at least 200 frames, to compute the offset map. The map is then regularly updated to mitigate, for instance, the emergence of new flickering hot pixels over time. When a new frame is transferred, the offset map is subtracted in real time and the resulting pixel amplitudes are compared to an event threshold where all pixels above it are considered as valid X-ray events. The event threshold can be defined as a fixed value or pixel-dependent using the initial frames of the sequence. In practice, we use a fixed event threshold value for all pixels tuned based on the best resulting spectral performance observed. The island around each valid event is extracted and finally saved. Given the small pixel size of the sensor (2.9~µm) and the likely small thickness of the device ($<$10~µm), we consider islands from $3\times3$ up to $7\times7$ pixels to ensure the extraction of X-ray events extend over multiple pixels. Then, a list of events considered as valid X-ray photons is regularly saved. Possible remaining hot pixels are rejected during the offline post-processing based on their anomalous count rate.
Although the frames can be partially reconstructed at a later time, one of the main limitations of this approach is that event extraction cannot be run again with other parameters such as different event threshold values or larger event island sizes. The algorithm was successfully run on a Raspberry Pi 4B and a Intel Core i7-13700K processor. The number of frames per second that can be processed by the CPU depends on many factors, such as the counting regime or the event island size. For our study, running at 10 frames per second allowed us to extract $\sim$7,000 events per second.

\section{Energy calibration}
\subsection{Calibration methods}
\label{subsect:calib_method}
In CMOS architecture, charge-voltage conversion and the first amplification stage are implemented directly at the pixel level. This means that each pixel has its own amplification circuit, likely generating non-uniform conversion gain between pixels. In addition to the Fano and electronic noise, small variations in the pixel gains can broaden the energy line and degrade the energy resolution of the sensor. To overcome this degradation, correction of the pixel-to-pixel gain dispersion can potentially improve the spectral performance. The X-ray calibration process involves illuminating the detector with a source emitting X-ray photons at known energies, ideally with several lines to increase calibration accuracy. The correlation between the observed spectrum (in ADU) and the known spectrum (in keV) can be measured using different approaches. A common method is the so-called ``peak fitting". It consists of fitting every line in the observed ADU spectrum with a Gaussian function to determine the line centroid and find the relation between these centers and the expected incident photon energy. A linear relation usually provides a reliable energy calibration, but a more complex relation (e.g., quadratic) can be employed to correct for non-linear readout electronic effects over a large energy range. Another possible approach is the ``Energy Calibration by Correlation" (ECC) \cite{maier2016a,maier2020a} initially employed for CdTe semiconductor detectors. The method relies on finding the maximum of correlation between a synthetic spectrum (template) of the incident source and the observed spectrum. Similarly to the peak fitting approach, the ECC method can be used to find linear or more complex relations. ECC offers multiple possible advantages over peak fitting. It allows a more robust and accurate calibration using lines and background as a whole, provides flexible ways of tracking the energy calibration evolution over time using a previous calibrated dataset, and has been shown to outperform peak fitting in the case of a low-statistics spectrum\cite{maier2016a}. The latter is particularly interesting for energy calibration of CMOS sensors, where a large number of pixel gains needs to be calculated, causing significant challenges in generating sufficient statistics for each pixel. Both methods were used in our study, and their results have been compared to determine the strength of each approach (Sec.~\ref{subsect:gain_meas}).

\begin{figure}[tp]
    \begin{center}
    \includegraphics[width=0.95\hsize]{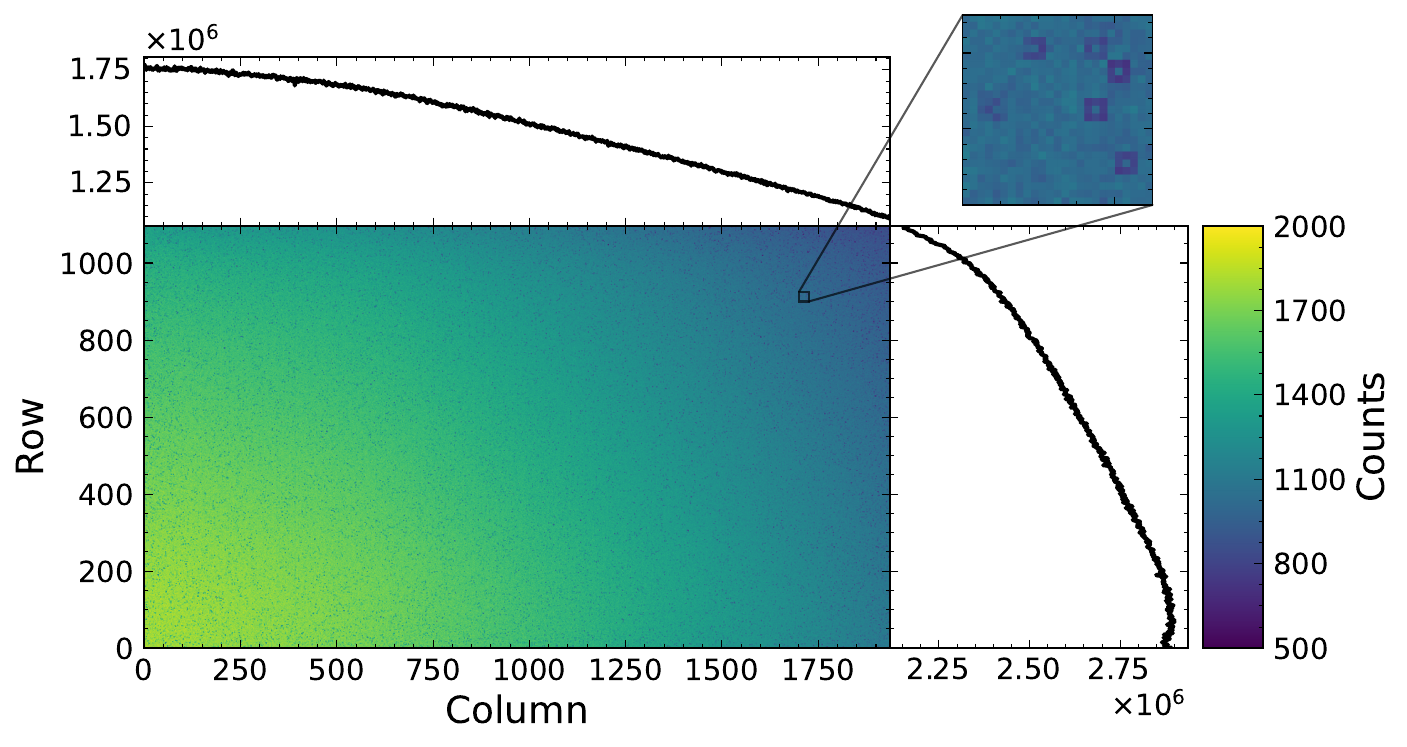} \\
    \includegraphics[width=0.95\hsize]{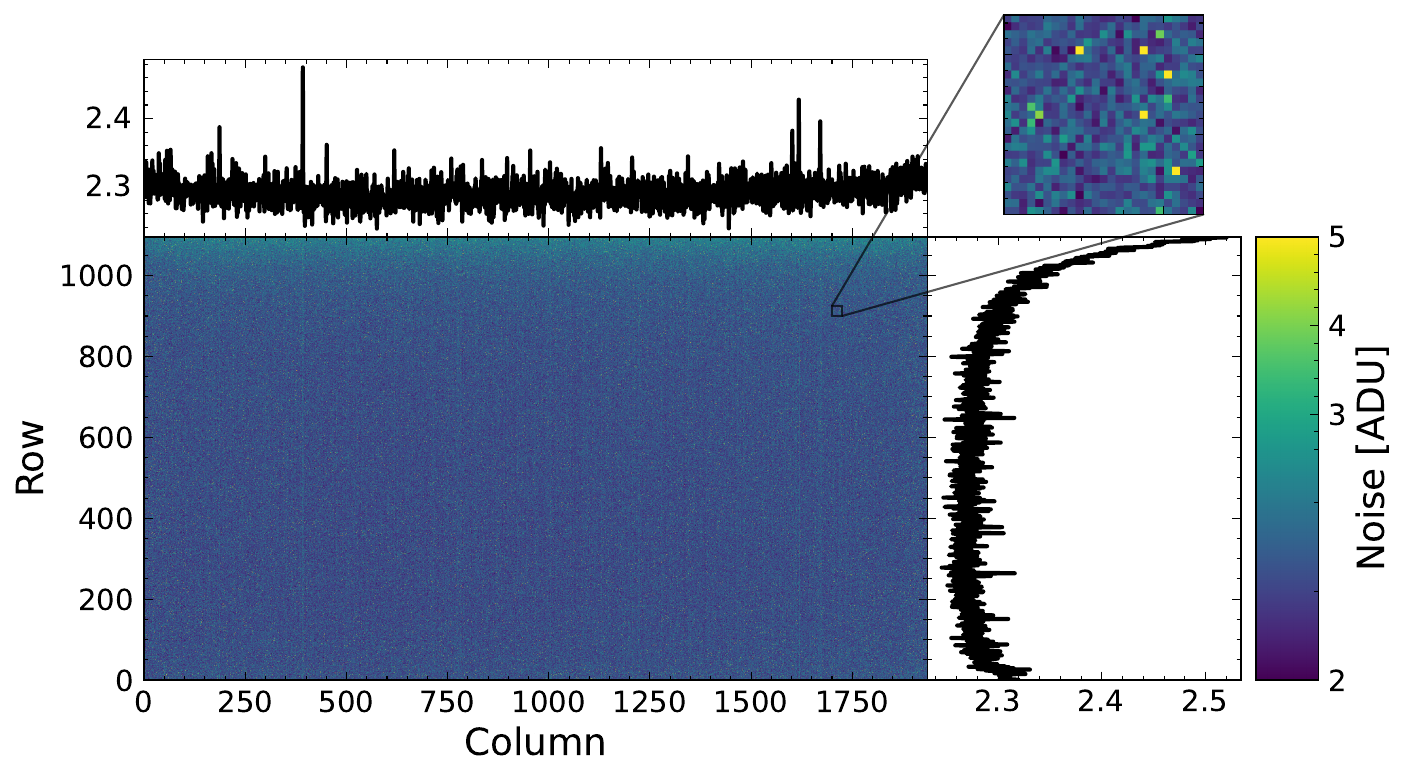}
    \end{center}
    \caption 
    { \label{fig:2D_counts}
    Top panel: Count map of single-pixel events obtained after 14 days of acquisition with the \fe\ source. The upper and right insets represent the sum of counts by column and row, respectively. The upper right inset shows a zoom of a small area of the sensor outlining the square patterns observed. Lower panel: Noise map derived from 300 frames at the beginning of one of the 10-min acquisition sequences. The upper and right insets show the average noise by column and row, respectively. The upper right inset shows a zoom in on a small region of the sensor, highlighting the pixels with higher noise than average producing the squared patterns observed on the count map.} 
\end{figure}

\subsection{Data collection for the energy calibration}
\label{subsect:data_processing}
The Sony IMX290 CMOS sensor is composed of 2.1 million pixels, and its precise energy calibration requires the measurement of a gain for each pixel. This requires generating a spectrum for each pixel, where each pixel spectrum is composed of single-pixel events to avoid mixing the gains of adjacent pixels. It also means collecting data under stable conditions to limit the contribution of external effects (e.g. temperature) on the gains measured.

To keep the acquisition time at a reasonable level, we illuminated the sensor with a bright \fe\ radioactive source ($\sim$38~MBq), which produces about 7,000 events/s on the detector. The camera was run at 10~fps to limit the pile-up and reduce the dark current to a negligible level for the operating conditions. To limit the data flow and save storage space, we performed real-time event extraction to capture only the relevant information, as described in Sec.~\ref{sect:data_processing}. The camera temperature was kept within 0.2°C throughout the acquisition using a TEC and a PID loop.

The \fe\ radioactive source produces two main energy lines at 5.90~keV and 6.49~keV. The fraction of single-pixel events obtained at 5.9~keV was measured to be $\sim$30\%. Previous work on the IMX290 attempted to measure the gain dispersion by dividing the sensor area into $5 \times 5$ sub-areas and measured a dispersion of 0.14\% \cite{tammes2020a}. Although this method can give a rough idea of the gain dispersion, it can still mix in small scale gain variations caused by mismatch variations of the pixel elements (e.g., transistors, photodiodes) during the manufacturing processes.

We performed \fe\ simulations to determine the optimal number of single-pixel events per pixel to accurately recover a given gain dispersion. We estimated that for a gain dispersion of $\sim$0.1\%, more than 1,000 events per pixel are needed with both calibration methods. A more precise estimate of the optimal number of individual events is discussed further in Sec.~\ref{subsect:optimal_num}. Based on these estimates and the count rate produced by the radioactive source, we targeted 14 consecutive days of acquisition to achieve $\sim$1,200 events per pixel.

Our acquisition sequence consists of two main steps. First, the offset map is calculated using 300~frames (30~s of data collection). The setup configuration prevents removal of the radioactive source during this acquisition time. We thus derived the offset map by applying a sigma clipping to mitigate the effect of X-ray events on the calculation. The second phase consists of collecting 6,000 consecutive frames (10~min of acquisition) and performing a real-time X-ray event extraction using the offset map as described in Sec.~\ref{sect:data_processing}. We repeat this sequence every 10~min for 14 days to track the evolution of the offsets and to ensure efficient real-time event extraction. When the camera is turned on, the temperature may temporarily fluctuate faster due to the additional heat produced by the readout electronics until a new equilibrium is reached. In order to have only datasets with a stable camera temperature, the first 1.5~h of datasets were removed from our post-processing analysis.

\subsection{Post-processing and per-pixel spectrum}
The data set collected over 14 days generated 1~TB of data. Processing such a large volume of data remains challenging with current laboratory computing resources. Thus, we performed the post-processing analysis on the MIT SuperCloud\cite{reuther2018a}, a supercomputing cluster at MIT that currently provides individual allocation of 384~CPUs.
The post-processing analysis was designed and optimized to minimize computational time, resulting in approximately one day of computation on 384~CPUs. Single-pixel events were first extracted from each 10-minute event list file and then redistributed to their respective pixels along the 14 days of acquisition. 
The top panel of Fig.~\ref{fig:2D_counts} shows the count map of the 2.1 million pixels from the measurement. We determined a median count of 1368 and observed a non-uniform distribution of counts across the sensor area. The lower left corner had a count of more than 2,000, while the upper right corner had a count of $\sim$500. This could be due to a misalignment of the \fe\ source with the sensor creating an inhomogeneous illumination pattern. The potential impact of this inhomogeneity on our analysis is discussed further in Sec.~\ref{subsect:gain_meas}. 
We also observed that a significant number of pixels exhibited a square pattern, where the central pixel has a higher count rate than the adjacent pixels (see the inset zoom region in Fig.~\ref{fig:2D_counts}). Similarly, when we extracted two-pixel events from the datasets, we noticed an opposite pattern trend in the count map, where the central pixel has a lower count rate than the adjacent pixels. These patterns appear to be associated with pixels with higher noise levels, as shown in the noise map in the bottom panel of Fig.~\ref{fig:2D_counts}. In this respect, pixels with higher noise levels cause the migration of single-pixel events into two-pixel events in adjacent pixels. This is a consequence of using the same split threshold for all pixels during the event extraction process. Implementing a pixel-dependent split threshold based on individual noise levels would mitigate this effect and result in a more homogeneous count map.

\begin{figure}[t]
    \begin{center}
    \includegraphics[width=0.7\hsize]{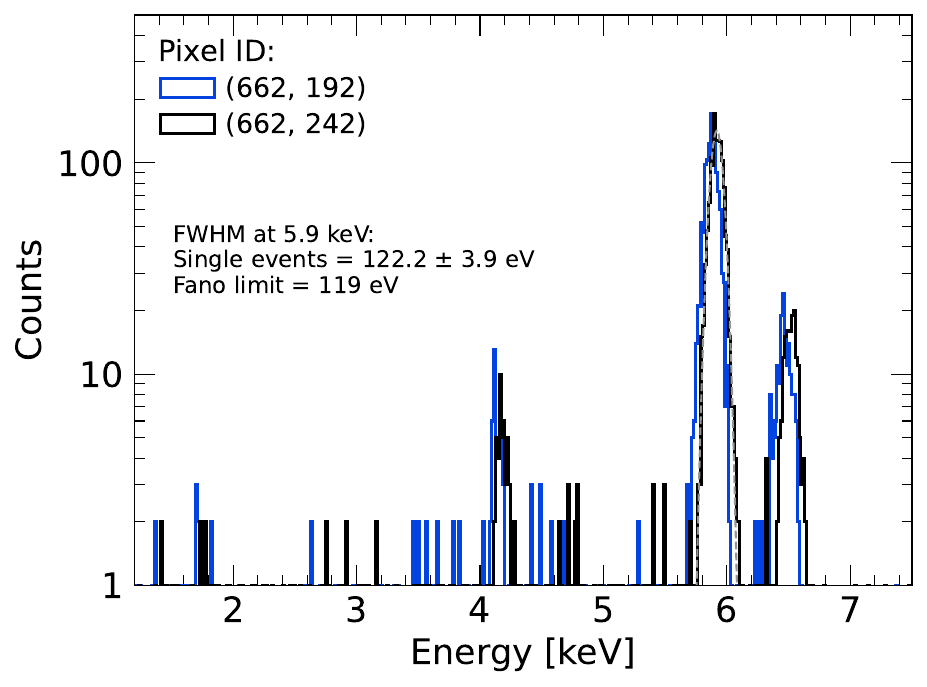}
    \end{center}
    \caption 
    {\label{fig:spec_fe55_2pix}
    \fe\ spectra of single-pixel events from the same column (\#662) but in different rows (\#192 vs. \#242) of the sensor area calibrated with the same gain value. The Mn-K$\alpha$ (5.90~keV) and Mn-K$\beta$ (6.49~keV) lines produced by the \fe\ radioactive source and the Si-K escape line of Mn-K$\alpha$ (4.16~keV) are detected in both spectra. A small horizontal shift is visible between the two spectra. The gray dashed line shows the best-fit Gaussian model used to measure the spectral energy resolution for the pixel ID (662, 242).} 
\end{figure}

We ended up with one spectrum for each of the 2.1 million pixels, made up entirely of single-pixel events. Figure~\ref{fig:spec_fe55_2pix} shows an example of two spectra from two pixels in the same column (\#662) but in different rows (\#192 vs. \#242). The two main emission lines produced by the radioactive source are clearly detected  at 5.90~keV (Mn-K$\alpha$) and 6.49~keV (Mn-K$\beta$). In addition, the Si-K escape line of Mn-K$\alpha$ (4.16~keV) is marginally detected. Both spectra were calibrated in energy with the same gain, and a small energy shift between them is noticeable, contributing to the broadening of the lines when combined. 
We measured a spectral resolution of 122~eV at 5.9~keV using a Gaussian fit, close to the intrinsic Fano limit at this temperature (119~eV) and consistent with a 2e- rms noise device.

\begin{figure}[t]
    \begin{center}
    \includegraphics[width=0.95\hsize]{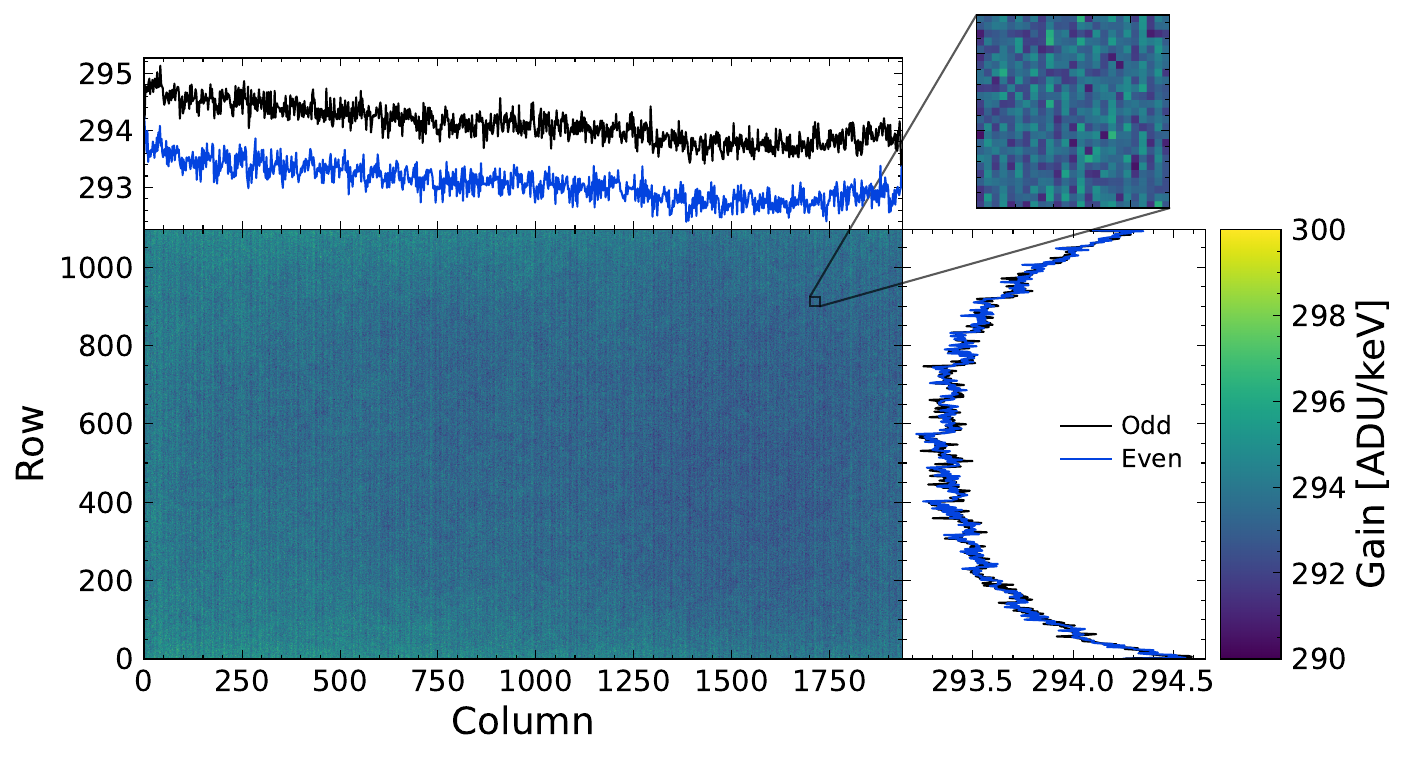}
    \end{center}
    \caption 
    { \label{fig:2D_gain}
    Gain map derived from \fe\ spectra of single-pixel events using the peak fitting method on the Mn-K$\alpha$ and Mn-K$\beta$ lines. The upper and right insets represent the average gain by column and row, respectively. Average gains for even (odd) columns and rows are indicated by blue (black) lines. The upper right inset shows a zoom of a small area of the sensor outlining the small scale gain variation.} 
\end{figure} 

\label{subsect:optimal_num}
\begin{figure}[t]
    \begin{center}
    \includegraphics[width=0.7\hsize]{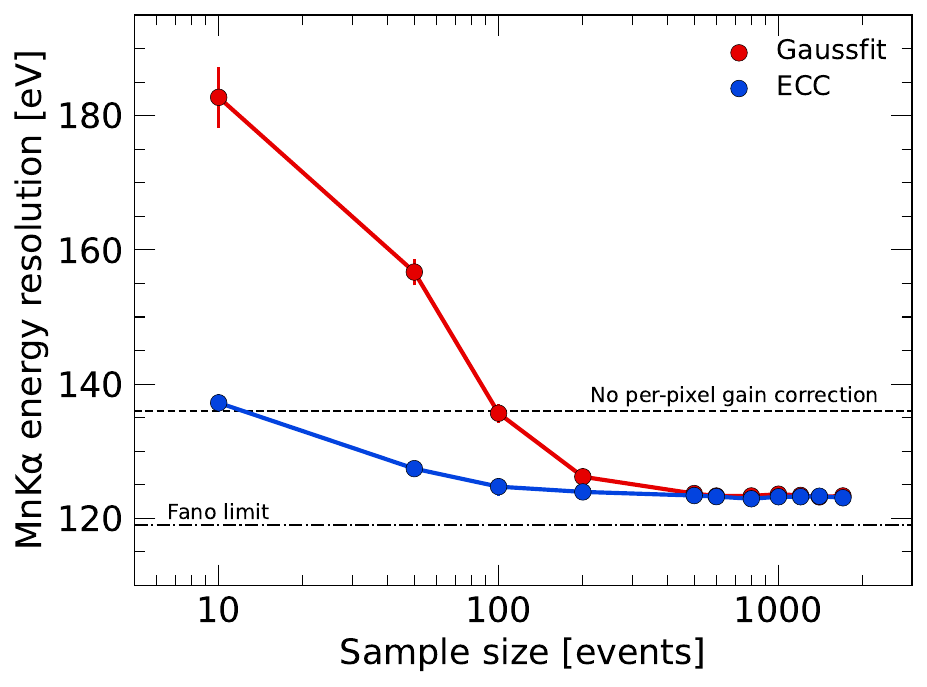}
    \end{center}
    \caption 
    { \label{fig:fe55_gainPerf}
    Energy resolution of single-pixel events at 5.9~keV (Mn-K$\alpha$) as a function of sample size. The energy resolution is measured using a Gaussian fit on a spectrum combining 10,000 pixels from the high-count region of Fig.~\ref{fig:2D_counts} and corrected for per-pixel gain variation using the peak fitting (red) and ECC (blue) methods. The dashed gray line represents the energy resolution assuming the same gain for all pixels. The Fano limit expected at room temperature is indicated by a dashed dotted line.} 
\end{figure} 
\begin{figure}[t]
    \begin{center}
    \includegraphics[width=0.49\hsize]{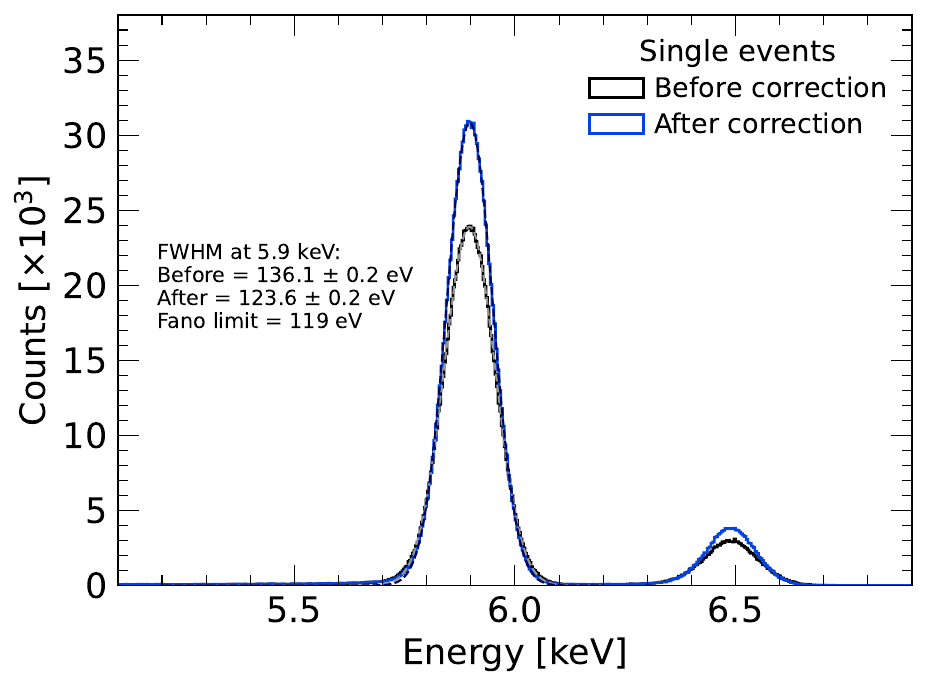}
    \includegraphics[width=0.49\hsize]{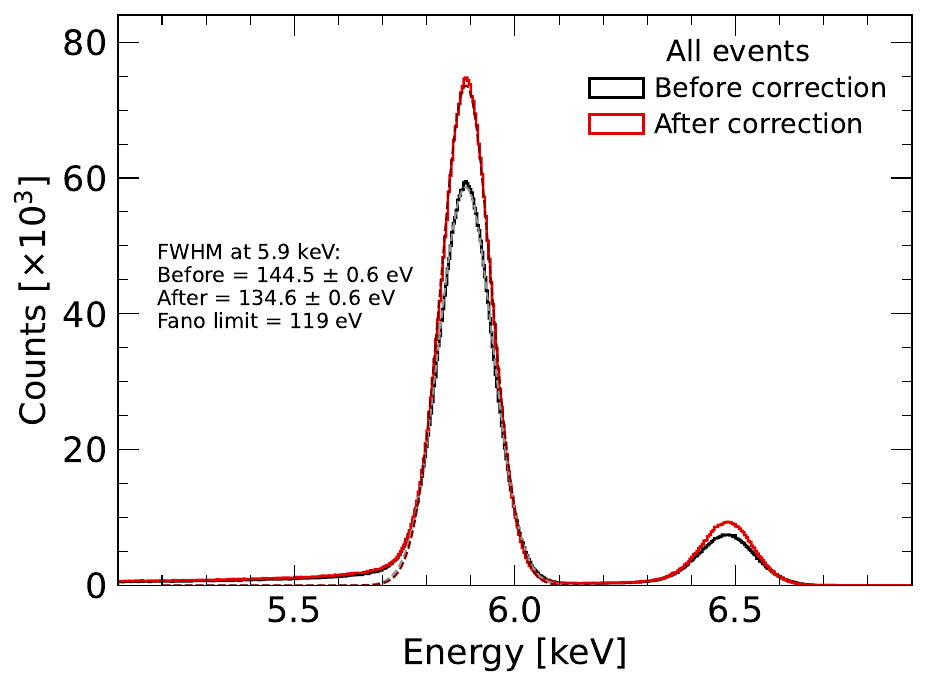}
    \end{center}
    \caption 
    { \label{fig:correction}
    Spectrum of single-pixel events (left) and all events (right) before per-pixel gain correction (in black) and after per-pixel gain correction (in blue or red). The gray, dark blue and dark red dashed lines show the best Gaussian fit used to derive the energy resolution.}
\end{figure} 

\subsection{Gain measurement and optimal counts per spectrum}
\label{subsect:gain_meas}
Two different calibration methods, peak fitting and ECC, were used to determine the gain for each pixel (see Sec.~\ref{subsect:calib_method}). We used the enhanced version of ECC with adaptive mesh refinement (AMR)\cite{maier2020a} to discretize the parameter space and reduce the computational time.

For both methods, we assumed a linear calibration in energy, expressed as $E = \mathrm{gain} \times PHA + \mathrm{offset}$, where PHA is the pulse height amplitude. Because the \fe\ source produces two emission lines within a limited energy range (0.5~keV), it can be challenging to constrain a small residual offset sometimes required to improve the energy calibration of some detectors. 
We first performed an energy calibration using the peak fitting method, considering two free parameters (gain and offset). The ratio of the offset to its error returned a median value of 1.2, suggesting that the residual offset, expected to be close to zero, cannot be well-constrained with the current data. This is supported by the real-time data processing, in which the offset map subtracted from the frames prior to extracting the X-ray events is updated every 10 minutes. This ensures that any systematic offset drift is minimized over time and that the residual offset should tend to zero. We therefore assumed an offset of 0 for the energy calibration in the rest of our analysis.

The gain map derived by the peak fitting method considering only one free parameter (gain) is shown in Fig.~\ref{fig:2D_gain}. The gain distribution shows a normal distribution with a mean of 294~ADU/keV and a 1-sigma dispersion of 0.4\%. This observed variation exceeds that reported in the literature using $5 \times 5$ binning regions (0.14\%) \cite{tammes2020a}, suggesting potentially significant interdevice variability or discrepancies due to the method used. 
The upper inset of Fig.~\ref{fig:2D_gain} represents the average gains for each column. A difference was observed between the average gains of odd and even columns. This difference is constant along the columns ($\sim$1~ADU/keV) and larger than the standard deviation of the average column gains.
This trend was not observed in the right inset of the 2D gain map (Fig.~\ref{fig:2D_gain}), where even and odd rows have consistent average gains along rows.
The feature observed for the average gain columns is thought to be due to the parallelism of the readout specific to CMOS technology, where the signal from an entire row can be read out simultaneously. In this design, each column has its own second stage signal amplifiers, correlated double sampling (CDS) and analog-to-digital converters (ADC). In addition, these components can be distributed on either side of the sensor area to save space, with even-column components on one side and odd-column components on the other.
No significant difference was observed between the gains measured by ECC and peak fitting methods, with $\widetilde{\textrm{gain}}_{\textrm{ECC}} = 293.66 \pm 1.18$~ADU/keV and $\widetilde{\textrm{gain}}_{\textrm{peak fitting}} = 293.59 \pm 1.18$~ADU/keV.

We investigated the performance and advantages of each method with respect to the number of single-pixel events in the spectrum (sample size). Due to the inhomogeneity observed in the count map in Fig.~\ref{fig:2D_counts}, which could bias the results, we performed this analysis by randomly selecting 10,000 pixels within the 90th to 95th percentile of the count distribution. This criterion yields an initial pixel sample of 104,737 pixels with counts between 1664 and 1720 photons. For each pixel, we generated a set of spectra with different sample size (10, 50, 100, 200, 500, ..., events) by randomly selecting events from their initial distribution.
For each sample size, we applied both ECC and peak fitting to the set of 10,000 spectra and measured their gains. The energy resolution of the Mn-K$\alpha$ line was then determined by combining the 10,000 spectra into a single spectrum after correcting for individual gain variations. The evolution of spectral resolution as a function of sample size is shown in Fig.~\ref{fig:fe55_gainPerf}. Both methods converge to the same energy resolution when spectra are composed of more than $\sim$500 single-pixel events. At lower counts, the ECC method outperforms peak fitting, showing an improvement in spectral performance for statistics greater than 10 photons. In contrast, peak fitting requires a minimum of 100 photons to achieve results similar to using a single gain for all pixels.
Given a gain dispersion of 0.4\%, this analysis provides guidance on the optimal statistics required to accurately correct for the per-pixel gain dispersion of the sensors. Since all pixels in our data set have more than 492 single-pixel events, it is unlikely that the inhomogeneous count pattern observed in the count map (Fig.~\ref{fig:2D_counts}) has affected our per-pixel gain correction.

\section{X-ray energy resolution after per-pixel gain correction}
\subsection{Per-pixel gain correction}
Correcting the per-pixel variation of CMOS sensors is expected to improve the spectral performance of the device. To validate our gain measurement, we corrected the gain dispersion using the gain map visible in Fig.~\ref{fig:2D_gain} on a 10-min data set, corresponding to about 4~million photons detected. 
The combined spectrum of single-pixel and all events before and after the gain correction is shown in Fig.~\ref{fig:correction}. As expected, the energy resolution is improved after correcting the per-pixel gain variation from $136.1\pm0.2$~eV to $123.6\pm0.2$~eV ($\sim$12~eV) for single-pixel events and from $144.5\pm0.6$~eV to $136.1\pm0.6$~eV ($\sim$10~eV) for all events spectra.

The improvement observed after the per-pixel gain correction confirms our ability to measure and correct the gain variation. The spectral performance of single-pixel events is approaching the theoretical limit of silicon detectors of 119~eV at room temperature (300~K) for an electron–hole pair creation energy of 3.67~eV and a Fano factor of 0.118\cite{lowe2007a}. The difference observed for single-pixel events compared to the Fano limit could be mainly attributed to the electronic noise of $\sim$2e- rms measured for this device. In addition, the electronic noise might increase the observed difference between single-pixel and all events to some extent by the fraction of energy loss below the split threshold, known as the charge sharing effect, which depends on the detector geometry and multiplicities. This effect can be mitigated empirically or by Monte Carlo simulations\cite{schneider2023a}, but was not corrected for in this analysis. Additional broadening of the lines compared to the Fano limit could result from the use of the same split threshold for the entire sensor area, producing the square patterns observed in Fig.~\ref{fig:2D_counts} and the overestimation of the energy of some events.

\subsection{Temperature dependence of gains}
\label{subsect:gaintemperature}
\begin{figure}[t]
    \begin{center}
    \includegraphics[width=0.49\hsize]{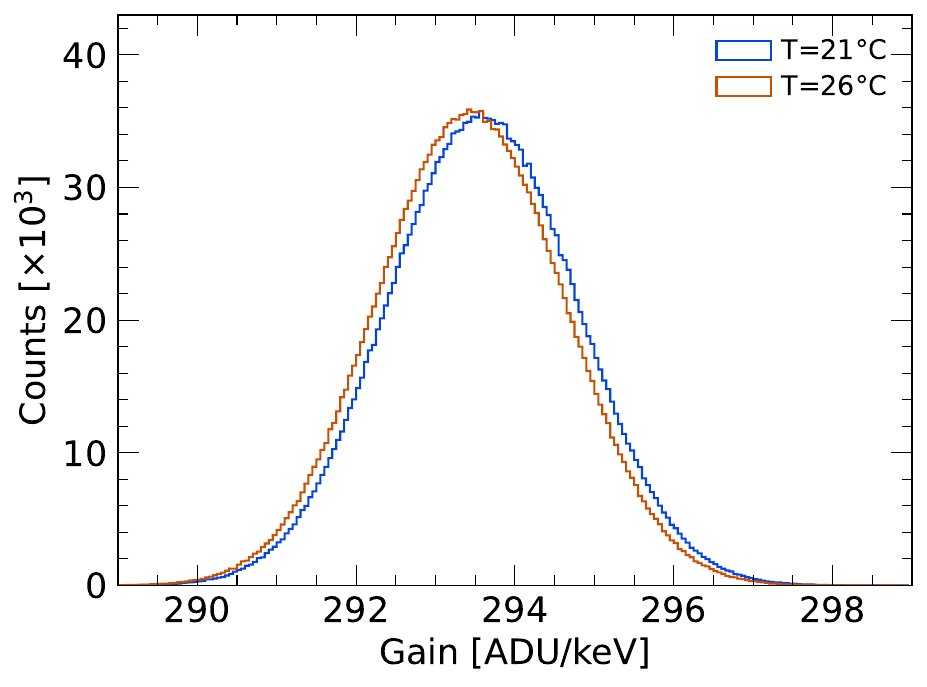}
    \includegraphics[width=0.49\hsize]{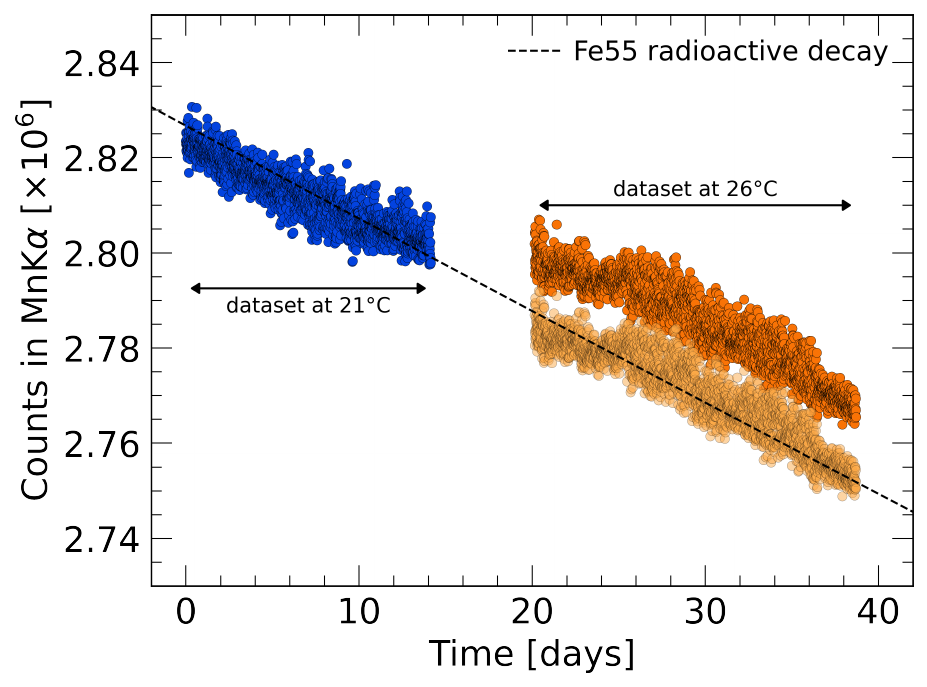}
    \end{center}
    \caption 
    { \label{fig:stability}
    Left panel: Histogram of the 2.1 million gains at 21°C (blue) and 26°C (orange). Right panel: Total number of counts per 10-minute file from the Mn-K$\alpha$ line ($5.7 < E < 6.1$~keV) as a function of time at 21°C (blue) and 26°C (orange). The dashed black line represents the expected radioactive decay rate of \fe\ during the observation period. The lighter orange circles indicate count rates at 26°C, reduced by a constant offset of 0.5\% to obtain a \fe\ decay rate consistent with the one measured on the 21°C data set. The exact cause for this offset remains to be explored in future work.} 
\end{figure}

After a pause of 6 days, we acquired a second data set under the same configuration and operating conditions except that the camera temperature was increased from 21°C to 26°C. The objective was to investigate the temperature dependence of the gain, in particular the per-pixel dispersion. The left panel of Fig.~\ref{fig:stability} shows the comparison of the gain distribution at the two temperatures. The gain dispersion at both temperatures was consistent at 0.4\%, but the distribution showed a median shift of 0.15 ADU/keV (0.05\%) when the temperature was raised by 5°C. At first order, this suggests that all pixel gains are affected in a similar way as the temperature changes, and that the gain dispersion remains constant within the temperature variations observed during data acquisition ($\lesssim$0.2°C). In a broader context, measuring the per-pixel gain dispersion at one temperature may offer the possibility of mitigating the gain dispersion at another temperature, at least within a certain temperature range. Additional data sets over a broader temperature range are needed to confirm the temperature dependence of the gains.

\subsection{Device stability over time}
Our long acquisition data sets allow us to measure the stability of the device over several days at a high count rate ($\sim$7,000 counts per second). 
In the right panel of the Fig.~\ref{fig:stability}, the total number of counts per 10~min file from the Mn-K$\alpha$ ($5.7 < \mathrm{E} < 6.1$~keV) is shown for both temperatures. Hot pixels have been removed from each file and the gain temperature dependence observed in the left panel of Fig.~\ref{fig:stability} has been corrected. 
We observed that at both temperatures the count rate in the Mn-K$\alpha$ line decreases progressively with time. The measured decay is consistent with the expected radioactive decay rate of \fe\ (half-life decay of 2.737 years) over this period. However, we noticed an additional constant offset of $\sim$0.5\% in the counts to the expected trend at 26°C. By removing this offset, we can obtain a decay consistent with the data set acquired at 21°C. The exact reasons for this offset will be the subject of future work. 

All pixels with an abnormal count rate (more than 10 counts per 10~min) were considered as hot pixels. The number of hot pixels was very stable over time for both temperatures, with a median number of hot pixels of $39 \pm 6$ at 21°C and $44 \pm 6$ at 26°C. At 26°C, we also saved the offset and noise map every 10 minutes to track their evolution over time. We found that the noise was very stable, with an equivalent electronic noise of $2.10^{+0.05}_{-0.01}$~e-~rms. The offset level was configured to a value of 200 ADU, and we measured a value of $199.76^{+0.60}_{-0.63}$~ADU, corresponding to a fluctuation of about 1~e- over 16 days. Our measurements demonstrated the excellent stability in terms of readout noise, hot pixels and dark current level of the camera over more than 30~days of data acquisition. 

\section{Conclusion}
We investigated the per-pixel gain calibration of the Sony IMX290 CMOS sensor composed of more than 2.1 million pixels. We collected X-ray photons from a \fe\ radioactive source under stable conditions over a period of two weeks at room temperature (21°C), in order to obtain more than 500 single-pixel events per pixel and generate a single-event spectrum per pixel.
We used two energy calibration methods (peak fitting and ECC) to measure the gain of all pixels, and investigated the possible advantages of each method. ECC outperformed peak fitting in the case of low statistics and offers a promising prospect for per-pixel gain correction of CMOS sensors.
Our gain measurements showed good overall gain homogeneity and a gain dispersion of 0.4\% for the entire sensor area. However, a constant difference along the columns of approximately 1 ADU/eV was noted between the average gains of the odd and even columns. Once corrected for per-pixel gain dispersion, we confirmed that the spectral performance was improved by $\gtrsim$10~eV at 5.9~keV. The spectral performance of single-pixel events was found close to the Fano limit at room temperature as expected for a 2e- rms readout noise device and demonstrated the impressive spectral performance of the sensor relative to its low cost. 
We repeated the measurement at a temperature +5°C higher (26°C) and observed a small shift (0.05\%) in the central value of the gain distribution, while the gain dispersion remained stable at 0.4\%. 
Furthermore, both data sets confirmed the excellent stability of the device with regard to noise, hot pixels and offsets over 30 days of acquisition under a count rate of more than 7,000 counts per second.
Further work on the Sony IMX290 CMOS sensor is needed to confirm its potential use in X-ray astronomy, such as measuring its X-ray quantum efficiency. 

\subsection*{Disclosures}
The authors declare no relevant financial interests or potential conflicts of interest related to this manuscript.

\subsection* {Code, Data, and Materials Availability} 
The data and codes utilized in this study are not publicly available and are the property of the Massachusetts Institute of Technology. They can be requested from the authors at bschn@mit.edu. The Energy Calibration by Correlation (ECC) code is the property of the Commissariat à l'Energie Atomique et aux Energies Alternatives (CEA).

\subsection* {Acknowledgments}
The authors acknowledge the MIT SuperCloud and Lincoln Laboratory Supercomputing Center for providing (HPC, database, consultation) resources that have contributed to the research results reported within this paper. They also acknowledge support from the MIT Kavli Institute's Research Investment Fund. 


\bibliographystyle{spiejour}   

\end{document}